\documentclass[sigconf]{acmart}
\usepackage{graphicx}
\AtBeginDocument{%
  \providecommand\BibTeX{{%
    \normalfont B\kern-0.5em{\scshape i\kern-0.25em b}\kern-0.8em\TeX}}}

\setcopyright{acmcopyright}
\copyrightyear{2024}
\acmYear{2024}
\acmDOI{XXXXXXX.XXXXXXX}

\acmConference[ACM conference]{ACM conference}{2024}{}
\acmPrice{15.00}
\acmISBN{978-1-4503-XXXX-X/18/06}

\begin{document}

\title{Turning Text and Imagery into Captivating Visual Video
}

\author{Mingming Wang, Elijah Miller}
\affiliation{%
  \institution{USC,wangmm913@gmail.com}
  \country{}
  }

\renewcommand{\shortauthors}{Mingming Wang}

\begin{abstract}

The ability to visualize a structure from multiple perspectives is crucial for comprehensive planning and presentation. This paper introduces an advanced application of generative models, akin to Stable Video Diffusion, tailored for architectural visualization. We explore the potential of these models to create consistent multi-perspective videos of buildings from single images and to generate design videos directly from textual descriptions. The proposed method enhances the design process by offering rapid prototyping, cost and time efficiency, and an enriched creative space for architects and designers. By harnessing the power of AI, our approach not only accelerates the visualization of architectural concepts but also enables a more interactive and immersive experience for clients and stakeholders. This advancement in architectural visualization represents a significant leap forward, allowing for a deeper exploration of design possibilities and a more effective communication of complex architectural ideas.

\end{abstract}



\maketitle

\section{Introduction}

The realm of architectural design~\cite{groat2013architectural,aliakseyeu2006computer,akin1996frames,lomas2007architectural,li2023archi} is inherently visual, with the ability to accurately and compellingly represent a structure from various perspectives being paramount to the success of any project. Traditional methods of architectural visualization, while effective, can be time-consuming and resource-intensive, often requiring specialized skills and extensive post-production work. With the advent of artificial intelligence and the development of advanced generative models, a new horizon in architectural visualization has emerged.

Diffusion-based methods~\cite{rombach2022high,song2020improved,imagen,improvedddpm,ho2020denoising,song2020score} are favored for their controllability, realism, and diversity, with applications in computer vision tasks such as image editing \cite{brooks2023instructpix2pix,hertz2022prompt2prompt,li2023layerdiffusion,ruiz2023dreambooth,tumanyan2023plug} and dense prediction \cite{diffusiondet,li2023efficient,diffmatch}. They are also used in video synthesis \cite{singer2022make,miller2024enhanced,ho2022imagenvideo,videocontrolnet,videoLDM,wang2018video2video,magicvideo,vdm} and 3D generation \cite{magic3d,dreamfusion,raj2023dreambooth3d,text-to-3d,tang2023dreamgaussian,li2024generating,diffusion3d}. In this paper, we present an innovative approach to architectural visualization that leverages the capabilities of models akin to Stable Video Diffusion~\cite{blattmann2023stable} to generate multi-perspective videos of buildings from single images and to create design videos directly from textual descriptions. We focus on the application of these models in the context of architectural design, exploring how they can be adapted and utilized to enhance the visualization process for architects, designers, and stakeholders.

Our approach to image-based multi-perspective video generation involves a meticulous preprocessing stage where architectural images are analyzed to identify key structural elements and potential viewpoints. This is followed by a sophisticated diffusion process that expands on the principles of SVD, generating a sequence of videos that showcase the building from various angles while maintaining visual coherence and continuity across frames.

For text-to-video generation in architectural design, we employ a model trained to interpret and encode textual descriptions of architectural concepts into a latent space. This encoded information serves as a foundation for the model to synthesize videos that bring the described architectural designs to life, offering a powerful tool for conceptual exploration and communication.

We also introduce innovations in camera motion control within the generated videos, allowing for the manipulation of viewing angles and paths to simulate realistic walkthroughs of the architectural designs. This feature is particularly beneficial for virtual presentations and client demonstrations, where the ability to navigate through a design is crucial.

Throughout the paper, we highlight the applications and benefits of our approach, demonstrating its potential to revolutionize the field of architectural visualization. From rapid prototyping to virtual presentations, our method streamlines the design process, making high-quality architectural visualization accessible and efficient.

Furthermore, we discuss the challenges and future work in this domain, including the need for broader generalization across diverse architectural styles and the integration of these models into existing design workflows. We also consider the ethical implications and the importance of maintaining the creative integrity of human designers in the age of AI.
\begin{figure*}[t]
	\centering

	\includegraphics[width=0.93\linewidth]{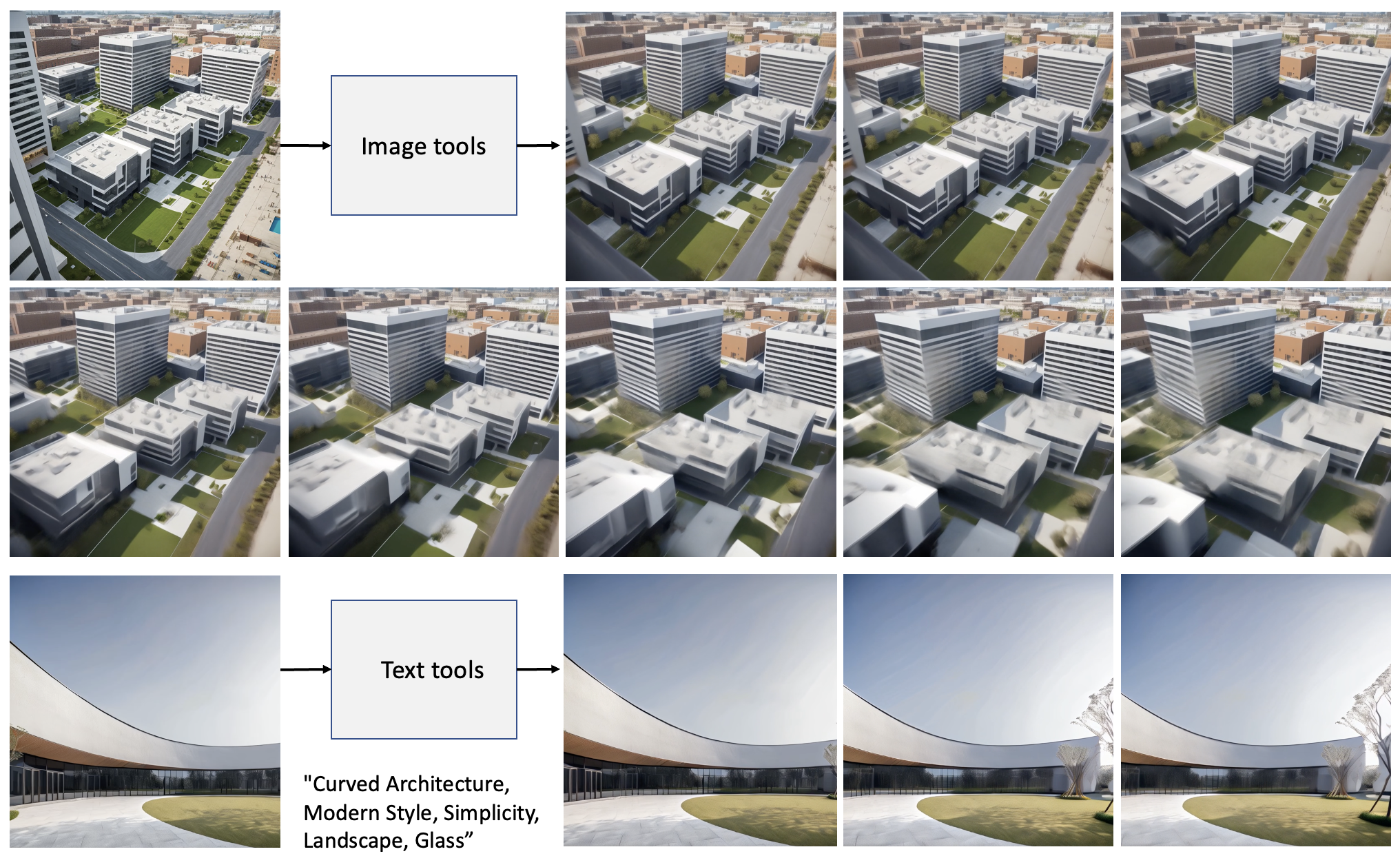}

	\caption{We show the capabilities of the generative model in creating multi-perspective architectural visualizations and transforming textual descriptions into dynamic video content.}
	\label{fig:tools}
\end{figure*}


\section{Application Details}
As shown in Fig. 1, our methodology is crafted on a foundation akin to Stable Video Diffusion (SVD), tailored and enhanced to address the unique demands of architectural visualization. The approach encompasses several pivotal components:

Image-Based Multi-Perspective Video Generation:

Preprocessing and Viewpoint Analysis: We initiate the process by meticulously preprocessing architectural images to extract essential structural elements and pinpoint optimal viewpoints.
Multi-Step Diffusion Process: Implementing a sophisticated multi-step diffusion process, our model crafts a sequence of videos that display the building from diverse vantage points, maintaining visual uniformity across frames.
Text-to-Video Generation for Architectural Design:

Text Encoding and Semantic Understanding: The model is adeptly trained to comprehend and encode textual depictions of architectural designs into a rich latent space.
Video Synthesis from Text: Capitalizing on this encoded data, the model synthesizes videos that vividly materialize the architectural designs as articulated in the text.
Model Architecture and Training:

UNet with Temporal Extensions: We harness a UNet architecture, thoughtfully expanded with temporal layers, to encapsulate the spatial and temporal intricacies of architectural structures.
Fine-Tuning for Architectural Data: The model is subjected to fine-tuning using a meticulously curated dataset of architectural images and videos, thereby enhancing its proficiency in generating nuanced building perspectives.
Innovations in Architectural Visualization:

Controlled Camera Motion: We introduce advanced modules that facilitate precise manipulation of camera motion within generated videos, offering granular control over viewing angles and trajectories.
Multi-View Consistency: The model steadfastly ensures consistency across various views, effectively simulating a seamless and realistic walkthrough of the architectural design.
Applications and Case Studies:

Rapid Prototyping: Designers can swiftly prototype architectural concepts through our video generation framework, streamlining the iteration process and fostering prompt feedback.
Virtual Presentations: The videos generated serve as an engaging medium for virtual presentations, allowing stakeholders to immerse themselves in the design from a multitude of perspectives.
Challenges and Future Work:

Generalization Across Architectural Styles: We confront the challenge of enhancing model generalization to encompass a wide array of architectural styles and their inherent complexities.
Integration with Design Workflows: We are committed to investigating more profound integrations with current architectural design and visualization workflows.
By fusing the principles of Stable Video Diffusion with the architectural design domain, our research introduces an innovative multi-perspective visualization approach, set to transform the conceptualization, development, and presentation paradigms for architects and designers.

\section{Conclusion}
In this paper, we introduce a application of diffusion-based models for architectural visualization, demonstrating the potential to revolutionize how architects and designers conceptualize and present their work. By harnessing the capabilities of Stable Video Diffusion-like models, we've presented a method that generates multi-perspective videos from single images and creates design videos directly from textual descriptions. This approach not only accelerates the design visualization process but also enhances the interactive and immersive experience for clients and stakeholders. As we look to the future, the integration of these models into standard design workflows and their ethical implications will be critical areas of focus, ensuring that the creative integrity of human designers remains at the forefront of the industry's evolution.

\bibliographystyle{ACM-Reference-Format}
\bibliography{sample-base}

\begin{thebibliography}{10}
\providecommand{\url}[1]{#1}
\csname url@samestyle\endcsname
\providecommand{\newblock}{\relax}
\providecommand{\bibinfo}[2]{#2}
\providecommand{\BIBentrySTDinterwordspacing}{\spaceskip=0pt\relax}
\providecommand{\BIBentryALTinterwordstretchfactor}{4}
\providecommand{\BIBentryALTinterwordspacing}{\spaceskip=\fontdimen2\font plus
\BIBentryALTinterwordstretchfactor\fontdimen3\font minus \fontdimen4\font\relax}
\providecommand{\BIBforeignlanguage}[2]{{%
\expandafter\ifx\csname l@#1\endcsname\relax
\typeout{** WARNING: IEEEtran.bst: No hyphenation pattern has been}%
\typeout{** loaded for the language `#1'. Using the pattern for}%
\typeout{** the default language instead.}%
\else
\language=\csname l@#1\endcsname
\fi
#2}}
\providecommand{\BIBdecl}{\relax}
\BIBdecl

\bibitem{groat2013architectural}
L.~N. Groat and D.~Wang, \emph{Architectural research methods}.\hskip 1em plus 0.5em minus 0.4em\relax John Wiley \& Sons, 2013.

\bibitem{aliakseyeu2006computer}
D.~Aliakseyeu, J.-B. Martens, and M.~Rauterberg, ``A computer support tool for the early stages of architectural design,'' \emph{Interacting with Computers}, vol.~18, no.~4, pp. 528--555, 2006.

\bibitem{akin1996frames}
{\"O}.~Akin and C.~Akin, ``Frames of reference in architectural design: analysing the hyperacclamation (aha-!),'' \emph{Design studies}, vol.~17, no.~4, pp. 341--361, 1996.

\bibitem{lomas2007architectural}
K.~J. Lomas, ``Architectural design of an advanced naturally ventilated building form,'' \emph{Energy and Buildings}, vol.~39, no.~2, pp. 166--181, 2007.

\bibitem{li2023archi}
P.~Li, B.~Li, and Z.~Li, ``{Sketch-to-Architecture: Generative AI-aided Architectural Design},'' in \emph{Pacific Graphics Short Papers and Posters}.\hskip 1em plus 0.5em minus 0.4em\relax The Eurographics Association, 2023.

\bibitem{rombach2022high}
R.~Rombach, A.~Blattmann, D.~Lorenz, P.~Esser, and B.~Ommer, ``High-resolution image synthesis with latent diffusion models,'' in \emph{CVPR}, 2022.

\bibitem{song2020improved}
Y.~Song and S.~Ermon, ``Improved techniques for training score-based generative models,'' in \emph{NeurIPS}, 2020.

\bibitem{imagen}
C.~Saharia, W.~Chan, S.~Saxena, L.~Li, J.~Whang, E.~L. Denton, K.~Ghasemipour, R.~Gontijo~Lopes, B.~Karagol~Ayan, T.~Salimans \emph{et~al.}, ``Photorealistic text-to-image diffusion models with deep language understanding,'' in \emph{NeurIPS}, 2022.

\bibitem{improvedddpm}
A.~Q. Nichol and P.~Dhariwal, ``Improved denoising diffusion probabilistic models,'' in \emph{ICML}, 2021.

\bibitem{ho2020denoising}
J.~Ho, A.~Jain, and P.~Abbeel, ``Denoising diffusion probabilistic models,'' in \emph{NeurIPS}, 2020.

\bibitem{song2020score}
Y.~Song, J.~Sohl-Dickstein, D.~P. Kingma, A.~Kumar, S.~Ermon, and B.~Poole, ``Score-based generative modeling through stochastic differential equations,'' in \emph{ICLR}, 2021.

\bibitem{brooks2023instructpix2pix}
T.~Brooks, A.~Holynski, and A.~A. Efros, ``Instructpix2pix: Learning to follow image editing instructions,'' in \emph{CVPR}, 2023.

\bibitem{hertz2022prompt2prompt}
A.~Hertz, R.~Mokady, J.~Tenenbaum, K.~Aberman, Y.~Pritch, and D.~Cohen-Or, ``Prompt-to-prompt image editing with cross attention control,'' in \emph{ICLR}, 2023.

\bibitem{li2023layerdiffusion}
P.~Li, Q.~Huang, Y.~Ding, and Z.~Li, ``Layerdiffusion: Layered controlled image editing with diffusion models,'' in \emph{SIGGRAPH Asia 2023 Technical Communications}, 2023, pp. 1--4.

\bibitem{ruiz2023dreambooth}
N.~Ruiz, Y.~Li, V.~Jampani, Y.~Pritch, M.~Rubinstein, and K.~Aberman, ``Dreambooth: Fine tuning text-to-image diffusion models for subject-driven generation,'' in \emph{CVPR}, 2023.

\bibitem{tumanyan2023plug}
N.~Tumanyan, M.~Geyer, S.~Bagon, and T.~Dekel, ``Plug-and-play diffusion features for text-driven image-to-image translation,'' in \emph{CVPR}, 2023.

\bibitem{diffusiondet}
S.~Chen, P.~Sun, Y.~Song, and P.~Luo, ``Diffusiondet: Diffusion model for object detection,'' in \emph{ICCV}, 2022.

\bibitem{li2023efficient}
P.~Li and Z.~Li, ``Efficient temporal denoising for improved depth map applications,'' in \emph{Proc. Int. Conf. Learn. Representations, Tiny papers}, 2023.

\bibitem{diffmatch}
J.~Nam, G.~Lee, S.~Kim, H.~Kim, H.~Cho, S.~Kim, and S.~Kim, ``Diffmatch: Diffusion model for dense matching,'' \emph{arXiv:2305.19094}, 2023.

\bibitem{singer2022make}
U.~Singer, A.~Polyak, T.~Hayes, X.~Yin, J.~An, S.~Zhang, Q.~Hu, H.~Yang, O.~Ashual, O.~Gafni \emph{et~al.}, ``Make-a-video: Text-to-video generation without text-video data,'' in \emph{ICLR}, 2023.

\bibitem{ho2022imagenvideo}
J.~Ho, W.~Chan, C.~Saharia, J.~Whang, R.~Gao, A.~Gritsenko, D.~P. Kingma, B.~Poole, M.~Norouzi, D.~J. Fleet \emph{et~al.}, ``Imagen video: High definition video generation with diffusion models,'' \emph{arXiv:2210.02303}, 2022.

\bibitem{videocontrolnet}
Z.~Hu and D.~Xu, ``Videocontrolnet: A motion-guided video-to-video translation framework by using diffusion model with controlnet,'' \emph{arXiv:2307.14073}, 2023.

\bibitem{videoLDM}
A.~Blattmann, R.~Rombach, H.~Ling, T.~Dockhorn, S.~W. Kim, S.~Fidler, and K.~Kreis, ``Align your latents: High-resolution video synthesis with latent diffusion models,'' in \emph{CVPR}, 2023.

\bibitem{wang2018video2video}
T.-C. Wang, M.-Y. Liu, J.-Y. Zhu, G.~Liu, A.~Tao, J.~Kautz, and B.~Catanzaro, ``Video-to-video synthesis,'' in \emph{NeurIPS}, 2018.

\bibitem{magicvideo}
D.~Zhou, W.~Wang, H.~Yan, W.~Lv, Y.~Zhu, and J.~Feng, ``Magicvideo: Efficient video generation with latent diffusion models,'' \emph{arXiv:2211.11018}, 2022.

\bibitem{vdm}
J.~Ho, T.~Salimans, A.~Gritsenko, W.~Chan, M.~Norouzi, and D.~J. Fleet, ``Video diffusion models,'' in \emph{NeurIPS}, 2022.

\bibitem{magic3d}
C.-H. Lin, J.~Gao, L.~Tang, T.~Takikawa, X.~Zeng, X.~Huang, K.~Kreis, S.~Fidler, M.-Y. Liu, and T.-Y. Lin, ``Magic3d: High-resolution text-to-3d content creation,'' in \emph{CVPR}, 2023.

\bibitem{dreamfusion}
B.~Poole, A.~Jain, J.~T. Barron, and B.~Mildenhall, ``Dreamfusion: Text-to-3d using 2d diffusion,'' \emph{arXiv preprint arXiv:2209.14988}, 2022.

\bibitem{raj2023dreambooth3d}
A.~Raj, S.~Kaza, B.~Poole, M.~Niemeyer, N.~Ruiz, B.~Mildenhall, S.~Zada, K.~Aberman, M.~Rubinstein, J.~Barron \emph{et~al.}, ``Dreambooth3d: Subject-driven text-to-3d generation,'' \emph{arXiv preprint arXiv:2303.13508}, 2023.

\bibitem{text-to-3d}
C.~Li, C.~Zhang, A.~Waghwase, L.-H. Lee, F.~Rameau, Y.~Yang, S.-H. Bae, and C.~S. Hong, ``Generative ai meets 3d: A survey on text-to-3d in aigc era,'' \emph{arXiv:2305.06131}, 2023.

\bibitem{tang2023dreamgaussian}
J.~Tang, J.~Ren, H.~Zhou, Z.~Liu, and G.~Zeng, ``Dreamgaussian: Generative gaussian splatting for efficient 3d content creation,'' \emph{arXiv preprint arXiv:2309.16653}, 2023.

\bibitem{li2024generating}
P.~Li and B.~Li, ``Generating daylight-driven architectural design via diffusion models,'' \emph{arXiv preprint arXiv:2404.13353}, 2024.

\bibitem{diffusion3d}
S.~Luo and W.~Hu, ``Diffusion probabilistic models for 3d point cloud generation,'' in \emph{CVPR}, 2021.

\bibitem{blattmann2023stable}
A.~Blattmann, T.~Dockhorn, S.~Kulal, D.~Mendelevitch, M.~Kilian, D.~Lorenz, Y.~Levi, Z.~English, V.~Voleti, A.~Letts \emph{et~al.}, ``Stable video diffusion: Scaling latent video diffusion models to large datasets,'' \emph{arXiv preprint arXiv:2311.15127}, 2023.

\end{thebibliography}

\end{document}